\documentclass[np2,twoside,twocolumn,showkeys]{revtex4}
\usepackage{kotex}
\usepackage{appendix}
\usepackage{amssymb}
\usepackage{amsmath}
\usepackage{graphicx}
\usepackage{epsfig}
\usepackage{color}
\usepackage{bm,multirow}
\usepackage{hyperref}
\usepackage{ulem}
\usepackage{siunitx}
\usepackage{verbatim}

\hypersetup{
colorlinks=true,
linkcolor=blue,
filecolor=magenta}
\usepackage{array} 

\begin{document}

\title[Distribution of Physics Departments]{Statistical Analysis on Scale and Regional Distribution of Undergraduate Physics Programs in Korean Universities}

\author{Gahyoun \surname{Gim}} 
\affiliation{Department of Physics, Gyeongsang National University, Jinju 52828, Korea}

\author{Sang Hoon \surname{Lee}} 
\affiliation{Department of Physics, Gyeongsang National University, Jinju 52828, Korea}

\date[]{Submitted \today}

\begin{abstract}
We report on the temporal changes in undergraduate-level physics programs at Korean universities from 1915 to 2023 by analyzing data on physics-related departments and their students using basic statistics and the scaling theory of statistical physics. Our analysis reveals that the number of departments peaked around the turn of the 21st century, and it has been steadily decreasing ever since, with particularly severe declines in private universities located outside the capital region. Besides the change in the overall numbers, we also show the change in the self-identity of physics-related departments reflected in department names, which reveals a recent trend of emphasizing more application-side such as semiconductors and data. As a sophisticated measure to quantify regional imbalances relative to the population eligible for higher education, we present scaling exponents from the scaling theory, which shows a shift from sublinear to linear for departments and a shift from linear to superlinear for students. The result indicates the exacerbation of the regional imbalance of university-level physics education in Korea. 
\end{abstract}
\keywords{Social Physics, Higher Education, Physics Department, Scaling Relation, Education Disparity}

\maketitle

\title[]{대한민국 대학의 물리학 학부과정 학과 및 학생 수의 증감과 지역적 분포 분석}

\author{김가현}
\email{dotchgahyoun@gnu.ac.kr}
\affiliation{경상국립대학교 물리학과, 진주 52828, 대한민국}
\author{이상훈}
\email[교신저자: ]{lshlj82@gnu.ac.kr} 
\affiliation{경상국립대학교 물리학과, 진주 52828, 대한민국}

\date[]{\today~투고}
\begin{abstract}
1915년부터 2023년까지 물리학 학부 과정을 운영한 대한민국 대학교의 물리학 관련 학과 데이터를 바탕으로, 기초적인 통계자료와 통계물리학의 스케일링 이론을 통해 물리학과 및 소속 학과 학생에 대한 지역별 변화를 살펴보았다. 물리학 관련 학과의 증감과 소속 학생의 증감 그리고 학과 명칭 변화에 대한 분석 결과 2000년 무렵 학과 수가 정점을 찍은 후 지속해서 감소가 일어나고 있음을 확인하였으며, 특히 지역 소재의 사립대학교를 중심으로 그러한 감소 경향이 심각하다는 것을 밝혔다. 학령인구 대비 지역별 분포에서 불균형 정도를 나타내는 거듭제곱 지수를 분석하였을 때, 학과의 거듭제곱 지수는 준선형에서 선형으로 나타났지만, 학생의 거듭제곱 지수는 선형에서 초선형으로 불균형한 정도가 더 크게 나타났다. 
\end{abstract}

\maketitlekorean
\setcounter{footnote}{0}

\section{서론}
최근 대한민국 사회에서 학령 인구 감소와 수도권 대학 선호도 증가 등 고등교육 관련 사회현상이 주목받고 있다\cite{studecline}. 이에 대응하여 많은 대학이 학과 통폐합을 진행하거나 신입생 유치를 목적으로 기초학문 학과를 첨단기술 분야로 전환하는 전략을 채택하고 있다. 이러한 흐름 속에서, `물리학과의 위기'라는 표현이 낯설지 않을 만큼 물리학 전공 학과들도 예외가 아니다. 예컨대 교육부의 대학 교육 편제 단위 표준 분류체계의 ``표준분류소계열 -- 물리학''으로 학부과정을 운영하던 학과들 중 2021학년 신입생 대상으로 경기대학교 및 동아대학교가 다른 계열로 전환하여 모집을 수행하였으며, 2023학년 신입생 대상 국립안동대학교 및 울산대학교가 계열 전환 후 신입생 모집, 그리고 같은 해 상지대학교가 신입생모집을 중단하였다.
앞서 언급한 기간동안 계열전환과 더불어 강릉원주대학교의 ``물리학과''가 ``수학물리학부 물리$\cdot$에너지전공''으로 기초과학계열 학과와의 학부단위의 통합이 이루어졌으며
국립공주대학교가 ``데이터정보물리학과''로, 국립목포대학교가 ``반도체응용물리학과''로 ``물리학과''에서 관련 첨단기술 분야를 포함하도록 학과 명칭을 변경했다. 또한, 2024학년도 기준으로 국립대학 소속의 ``물리학과'' 중 국립창원대학교 및 강원대학교가 ``반도체물리학과''로, 그리고 국립군산대학교가 ``첨단기술학부 반도체물리전공''으로 학과 명칭을 변경하였다\cite{Univalram}.

본 연구에서는, 이러한 국내 대학교의 물리학 관련 학과 및 학생 수 증감에 대한 기본적인 통계를 제시하고, 지역별 분포에 있어서의 편차와 그 의미를 통계물리의 방법론을 통해 정량적으로 분석할 것이다. 특히 신입생 모집 중단과 학과명칭의 변화의 주된 기인을 추적하기 위하여 소속 학교의 설립 주체, 지역, 종합대학 또는 과학기술특성화 대학 여부 등으로 구분하여 통계를 살펴볼 것이며, 지역별 학령인구 대비 물리학 관련 학과의 분포를 스케일링 이론 (scaling theory) 기반의 분석을 통해 체계적으로 정량화하고 그 의미를 도출할 것이다. 물론 그러한 분석을 위해서는 물리학 학부 과정에 대한 정확한 데이터가 필수적이나, 현재 국내 물리학 학부 과정을 운영하는 학과 및 소속 학생의 현황을 100\% 정확하게 파악하는 데는 다소의 어려움이 있다. 교육부가 매년 발간하는 `교육통계연보'에서 ``화학'', ``수학''과 같이 독립적으로 집계되거나 ``지구$\cdot$지리학'' ``천문$\cdot$기상학''와 같이 인접분야와 통계처리 되는 다른 기초 과학 계열과 달리 ``물리$\cdot$과학''으로 상위 분류와 함께 포괄적으로 집계하므로, 학과 및 학생 증감을 명확하게 알 수 없다. 한국물리학회가 제공하는 ``대학 물리학과''\cite{kpsUniv}에 기재된 정보는 비정기적으로 일부만 개정되어 현재 신입생 모집을 중단한 학과 정보가 포함되어 있는 문제가 있다. 본 연구에서는, 교육부와 한국물리학회의 기준을 바탕으로 한 데이터 수집 및 현실을 최대한 반영한 정제를 통해 정리한 데이터를 기반으로, 물리학 학부 과정을 제공하는 학과와 그에 소속된 학생 수에 대하여 매년 달라지는 값을 정리하여 분석하였다.

그러한 전체적인 증감과 더불어, 통계물리학에서 개발된 인구밀도와 시설 밀도의 거듭제곱 관계를 통하여 스케일링 지수의 값에 따라 시설의 성향을 정량화하는 스케일링 이론\cite{newman,PNASScaling}에 기반한 연구를 수행하였다. 참고문헌\cite{PNASScaling}에 보고된 기존 연구에서는, 전체 인구 대비 고등교육기관을 대학 교육 편제단위 표준 분류체계에서 ``표준분류대계열''단위로 스케일링 또는 거듭제곱 지수(exponent)를 측정한 바 있다. 본 논문의 저자들이 참여한 선행 연구\cite{OurSaemulli}에서는 고등교육기관을 전문대학과 4년제 대학교로 구분하고 소속 학과들을 대계열별로 구분하여 스케일링 지수로 표현되는 학령인구 대비 지역별 분포의 특성을 정량화하였고, 시간에 따른 전체적인 지역별 편차가 심화되고 있으며 정부 부처가 정원을 조절하는 의약계열과 교육계열의 지역 안배 특수성이 스케일링 지수에도 반영됨을 확인하였다. 본 연구의 스케일링 분석 부분에서는, 선행 연구에서 기초자연과학 계열로 통합되어 있어 따로 분석하지는 못했던 물리학 계열 학과에 집중하여 고등교육 대상 연령 인구 밀도 대비 광역자치단체별(특별시$\cdot$광역시$\cdot$도) 학과 밀도와 학생 밀도의 스케일링 지수를 도출하였다. 

선행 연구\cite{OurSaemulli}에서 사용한 학령인구밀도와는 달리 본 연구에서 기준으로 삼은 인구 밀도는 국가통계포털 (KOrean Statistical Information System: KOSIS)\cite{kosis}의 지역별, 연도별 만 15세부터 24세의 인구 데이터를 기반으로 한다. 이는 지역의 고등교육 수요 및 고등교육 대상자 연령대의 학업 이수 목적의 소재지 변경 경향을 좀 더 직접적으로 반영할 수 있도록, 각 지역에서 고등교육을 수요하는 15세부터 17세의 연령의 인구 및 국가지표체계에 명시된 고등교육 대상인 만 18세부터 21세와 4년의 학부 교육 과정을 고려한 것이며, 실제로 학령인구밀도와 비교하는 것에 비해 회귀 분석의 결정계수가 높게 나타나는 것을 확인하였다. 최종적으로 지역별, 연도별 만 15세부터 24세의 인구밀도 데이터와 지역별 및 연도별로 물리학 학부 과정의 운영 학과 밀도, 재학생 밀도의 스케일링 관계에 대한 세부 데이터를 분석하여 밀도값의 스케일링 지수를 도출하고, 시간에 따른 변화를 분석하여 경향성을 정량화했다.

\section{연구 대상 및 데이터}
\label{sec:data}

\subsection{연구 대상}
체계적인 데이터 수집을 위해, 본 연구에서는 교육부의 대학 교육 편제 단위 표준 분류체계의 ``표준분류소계열 -- 물리학'' 해당여부와 한국물리학회의 자료를 기반으로 분석 대상 학과를 선정하였다. 특히 한국물리학회의 제공 자료에서는 `한국 물리학회 50년사 -- 한국물리학회 분야별 50년사 -- 분회'\cite{kps50th} 중 교육계열을 제외한 전 학과 및 `한국물리학회 -- 물리학정보 -- 유용한사이트 -- 대학 물리학과'의 정보를 기반으로 했다. 1915년부터 2023년까지, 매년 3월 1일부터 이듬해 2월 마지막 날 까지 기간을 기준으로 학과 명칭과 소속 단과대학의 정보를 수집하였다. 또한 연구대상의 소속 학교 및 단과대학 학칙 규정 자료집을 통해, 연도별로 학과 명칭, 소속 단과대학의 변경 여부 및 변경 일자에 관한 데이터를 취합했다. 본 연구에서 사용한 원본 데이터는 Appendix~\ref{appendix : data}에 공개하였으며, 수집한 데이터는 Appendix~\ref{appendix : exception}에 기재한 일부 특수 사례에 대한 예외 규정을 적용하여 처리하였다.

\subsection{데이터 처리 기준 및 특수사례}
학과 명칭 변경과 지속 운영 여부는 매년 3월 1일을 기준으로, 신입생이 입학하는 학과 명칭, 해당 학과가 소속된 단과대학, 그리고 해당 연도의 신입생 모집 여부를 통해 판별하였다. 학교의 소재 지역은, 학과 혹은 단과대학의 행정상 소재 지역을 기준으로 분류하였으며 이원화 캠퍼스 대상 학과의 경우 행정상 학교 소재 지역을 따랐다. 학교의 설립 주체 변경, 학교 명칭 변경, 학교 및 캠퍼스 통합에 따른 물리학과 학부과정 간의 통합, 행정상 소재 지역 변경 등의 변화하는 행정지표에 대해서 2023학년도 기준으로 집계 후 반영하였다. 단, 세종특별자치시 소재 학교는 행정구역 분리 이후에도 단일 행정구역으로서의 스케일링 분석이 의미가 없을 만큼 규모가 작기에, 편의상 충청남도 소재 학교로 집계하였다.
\label{countrule}

\subsection{데이터 요약}

\begin{table*}
\caption{The total of $81$ physics-related departments analyzed in our study are categorized based on the type of university founder (public or private), their administrative regions, and the year of their foundation (the first undergraduate admission). See Figs.~\ref{figure : Fig-1} and \ref{figure : Fig-2} for the graphic representation in the cases of the type of university founders and the region.}

\begin{tabular}{c|cl|r|r}
\hline
\hline
 & \multicolumn{2}{c|}{private} & 54 & 66.67\% \\ \cline{2-5} 
 & \multicolumn{1}{c|}{} & National & 20 & 24.69\% \\ \cline{3-5} 
 & \multicolumn{1}{c|}{} & \begin{tabular}[c]{@{}l@{}}Institution of Science and Technology\end{tabular} & 4 & 4.94\% \\ \cline{3-5} 
 & \multicolumn{1}{c|}{} & \begin{tabular}[c]{@{}l@{}}National University Cooperation\end{tabular} & 2 & 2.47\% \\ \cline{3-5} 
\multirow{-5}{*}{\begin{tabular}[c]{@{}c@{}}types of \\ university\\founders\\(as of 2023)\end{tabular}} & \multicolumn{1}{c|}{\multirow{-4}{*}{\begin{tabular}[c]{@{}c@{}}non-private\\(33.33\%)\end{tabular}}} & Public & 1 & 1.23\% \\ \hline
 & \multicolumn{1}{c|}{} & Seoul & 19 & 23.46\% \\ \cline{3-5} 
 & \multicolumn{1}{c|}{} & Gyeonggi-do & 9 & 11.11\% \\ \cline{3-5} 
 & \multicolumn{1}{c|}{\multirow{-3}{*}{\begin{tabular}[c]{@{}c@{}}located in the capital \\ region (37.04\%)\end{tabular}}} & Incheon & 2 & 2.47\% \\ \cline{2-5} 
 & \multicolumn{1}{c|}{} & Gyeongsangbuk-do & 7 & 8.64\% \\ \cline{3-5} 
 & \multicolumn{1}{c|}{} & Busan & 6 & 7.41\% \\ \cline{3-5} 
 & \multicolumn{1}{c|}{} & Gyeongsangbuk-do & 5 & 6.17\% \\ \cline{3-5} 
 & \multicolumn{1}{c|}{} & Jeollabuk-do & 5 & 6.17\% \\ \cline{3-5} 
 & \multicolumn{1}{c|}{} & Daejeon & 5 & 6.17\% \\ \cline{3-5} 
 & \multicolumn{1}{c|}{} & Chungcheongnam-do & 5 & 6.17\% \\ \cline{3-5} 
 & \multicolumn{1}{c|}{} & Gyeongsangnam-do & 4 & 4.94\% \\ \cline{3-5} 
 & \multicolumn{1}{c|}{} & Chungcheongbuk-do & 3 & 3.70\% \\ \cline{3-5} 
 & \multicolumn{1}{c|}{} & Gwangju & 3 & 3.70\% \\ \cline{3-5} 
 & \multicolumn{1}{c|}{} & Jeollanam-do & 3 & 3.70\% \\ \cline{3-5} 
 & \multicolumn{1}{c|}{} & Daegu & 2 & 2.47\% \\ \cline{3-5} 
 & \multicolumn{1}{c|}{} & Ulsan & 2 & 2.47\% \\ \cline{3-5} 
\multirow{-16}{*}{\begin{tabular}[c]{@{}c@{}} region \\(as of 2023)\end{tabular}} & \multicolumn{1}{c|}{\multirow{-13}{*}{\begin{tabular}[c]{@{}c@{}}located outside \\ the capital region\\(62.96\%)\end{tabular}}} & Jeju & 1 & 1.23\% \\ \hline
 & \multicolumn{2}{l|}{before 1950} & 7 & 8.64\% \\ \cline{2-5} 
 & \multicolumn{2}{l|}{1950--1959} & 14 & 17.28\% \\ \cline{2-5} 
 & \multicolumn{2}{l|}{1960--1969} & 3 & 3.70\% \\ \cline{2-5} 
 & \multicolumn{2}{l|}{1970--1979} & 7 & 8.64\% \\ \cline{2-5} 
 & \multicolumn{2}{l|}{1980--1989} & 37 & 45.68\% \\ \cline{2-5} 
 & \multicolumn{2}{l|}{1990--1999} & 10 & 12.35\% \\ \cline{2-5} 
\multirow{-7}{*}{\begin{tabular}[c]{@{}c@{}}the first year of \\ undergraduate-student \\ entrance \end{tabular}} & \multicolumn{2}{l|}{after 2000} & 3 & 3.70\% \\ \hline\hline
\end{tabular}
\label{table:classification}
\end{table*}

\begin{figure}
\includegraphics[width=0.49\textwidth]{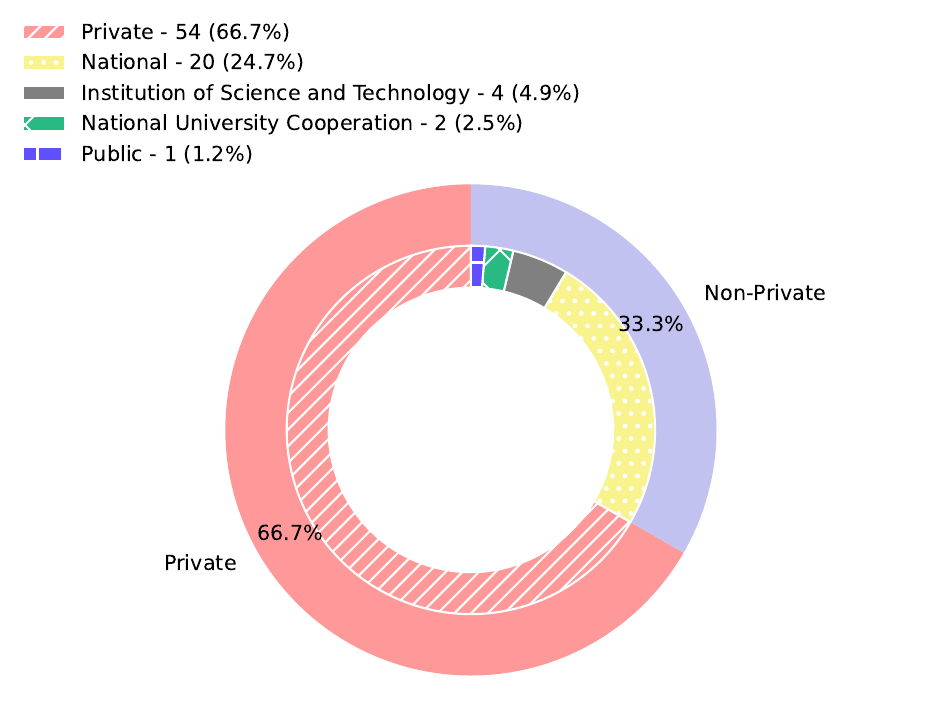}
\caption{The $81$ physics-related departments are largely categorized based on the type of university founder: public or non-private (the outer circular chart). The non-private category is further classified into national, institution of science and technology, national university cooperation, and public (the inner circular chart).
}
\label{figure : Fig-1}
\end{figure}


\begin{figure}
\includegraphics[width=0.49\textwidth]{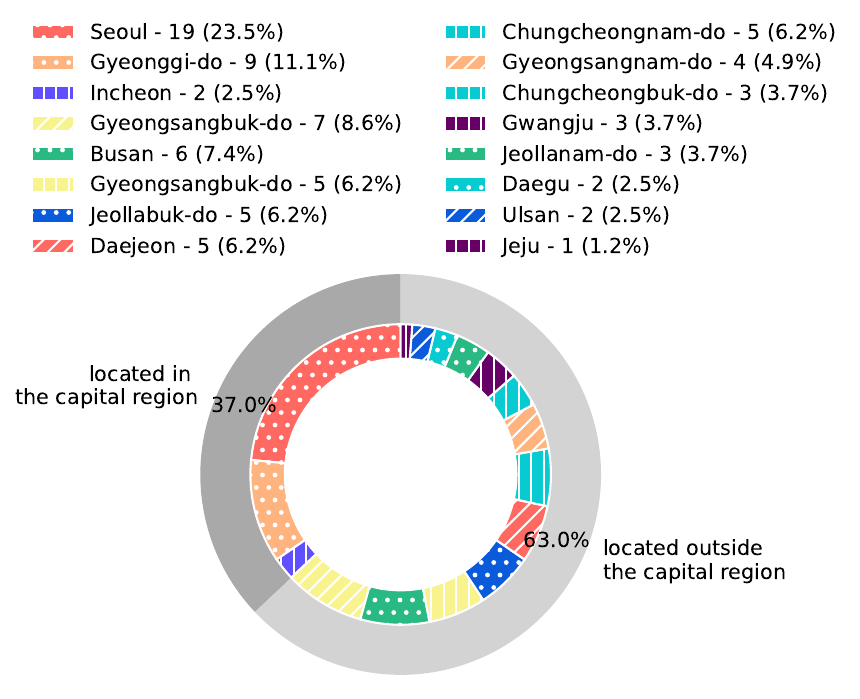}
\caption{The $81$ physics-related departments are categorized based on their location in the unit of administrative regions. The outer dark and light gray circular chart distinguishes the capital and non-capital regions, while the inner circular chart represents their respective sub-regions. For better visual inspection, we note that the regions in the inner pie chart are counterclockwise arranged in the same order as written in the legend (from the top left to the bottom right). 
}

\label{figure : Fig-2}
\end{figure}


Table~\ref{table:classification}, Fig.~\ref{figure : Fig-1}, Fig.~\ref{figure : Fig-2}\는 연구 대상인 81개 모든 학과를 설립 주체별, 지역별, (10년 단위로 구분한) 첫 학부 과정을 모집한 시기의 세 가지 범주로 분류한 데이터를 요약정리한 결과이다.
Figure~\ref{figure : Fig-1}의 도넛형 차트로 나타낸 것과 같이 소속 학교의 설립 주체로 구분하였을 때, 전체 집계 대상 81개의 학과 중 54개(66.67\%)의 학과가 사립대학에 소속되어 있는 것으로 나타났다. 사립대학 소속이 아닌 학교에서 개설한 물리학 학부 과정 학과 중 20개(24.69\%)가 국립대학에 소속되어 있으며, 2개(11.11\%)의 학과가 국립대법인 대학 소속, 4개(4.94\%)의 학과가 과학기술 특성화대학 (소위 ``IST'' 대학) 소속, 한 개(1.23\%)의 학과가 공립대학 소속으로 집계되었다.

마찬가지로 Fig.~\ref{figure : Fig-2}에서 나타낸 것과 같이 소속된 학교의 행정적 소재 지역으로 구분하였을 때, 30개의 학과(37.04\%)가 수도권 지역 대학에 소재하는 것으로 집계되었다. 특히 수도권 소재 학과 중, 서울 소재의 학교 소속 학과가 19개(23.46\%), 경기도 소재의 대학 소속 학과가 9개(11.11\%), 그리고 인천 소재 학교의 소속 학과가 2개(2.47\%)로 나타났다. 또한 비수도권 지역 대학 소속 학교는 총 51개(62.96\%)로 나타났으며 그 중 경상북도가 7개(8.64\%), 부산광역시는 6개(7.41\%), 경상북도, 전라북도, 대전광역시, 충청남도에 각 5개(6.17\%), 경상남도 4개(4.94\%), 광주광역시, 전라남도에 각 3개(3.70\%), 대구광역시 및 울산광역시는 각 2개(2.47\%), 마지막으로 제주특별자치도에 1개(1.23\%)의 학과가 대상 지역에 소재한 학교 소속으로 집계되었다.

학부 모집을 시작한 첫 연도를 10년 단위로 나누어 보았을 때, 1950년 이전에 7개(8.64\%)의 학부 과정 모집을 시작하였고, 1950년대에는 14개(17.28\%)의 학과가 개설되었다. 1960년대에는 3개(3.70\%), 1970년대에는 7개(8.64\%), 1980년대에는 37개(45.68\%)의 학과가 신설되었다. 1990년대에는 10개(12.35\%)의 학과가 학부 과정 모집을 시작하였고, 마지막으로 2000년 이후에는 3개(3.70\%)의 학교가 학부 모집을 시작한 것으로 확인되었다.

\def\msun{M_\odot}

\section{물리학 관련 학과 수의 시간적 증감}
\label{section : dept data}

\begin{figure*}
\includegraphics[width=0.9\textwidth ]{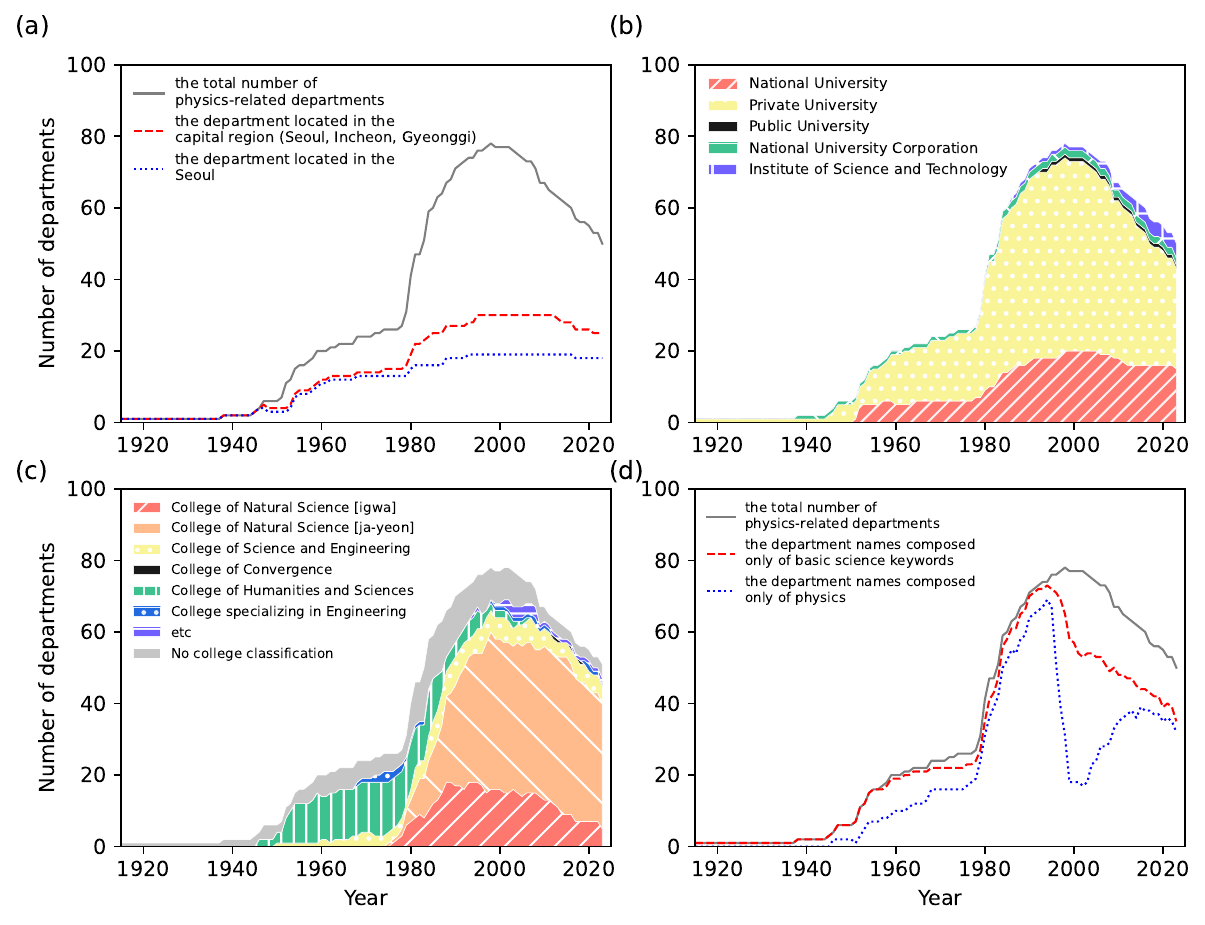} 
\caption{The temporal change of the number of physics-related departments, whose compositions are categorized by (a) the region, (b) the college to which each department belongs, and (c) the type of university founder. The panels (a) and (d) are line graphs representing the raw data for each category, while the panels (b) and (c) indicate the cumulative number stacked on top of one another (so the outermost boundary is identical to the total number of physics-related departments, as one can readily see). In panel (d), 
the red curve represents the number of physics-related departments whose names composed only of natural science keywords: basic, mathematics, chemistry, biology, earth science, astronomy, and physics (the intermediate-level classification, 
and the blue curve represents the number of physics departments whose names do not include any other modifiers than physics (the most strict classification). 
}

\label{figure : Fig-3}
\end{figure*}


집계 대상 학과 수의 시간에 따른 증감을 확인하기 위해 \ref{countrule}절에서 언급한 기준으로 학과 변화를 살펴보았다. 우선, 전 지역의 학과의 수, 서울 소재 학교 및 인천 경기를 포함한 수도권 소재 학교의 증감을 Fig.~\ref{figure : Fig-3}(a)에 나타냈다. 모든 광역자치단체 단위에 대한 변화는 Appendix의 Fig.~\ref{figure : Fig-appendix}에 누적그래프를 통하여 나타내었다. 1980년 이전까지는 서울에 위치한 학교 소속 학과들이 대상 기간동안 과반수 이상을 차지했으나, 1980년 이후 비수도권에 위치한 학과 수도 크게 증가한 것을 알 수 있다. 2000년부터는 수도권(주로 서울)에 있는 학과 수는 거의 유지되고 있는 반면, 물리학 관련 학과 집계 대상에서 제외된, 신입생을 모집하지 않거나 계열을 전환한 학과 대다수는 비수도권에 소재한 학과로 나타났다. 

2023학년도를 기준으로 한 설립 주체별로 구분하여 물리학 학부 과정 학과의 증감을 나타낸 것이 Fig.~\ref{figure : Fig-3}(b)이다. 비교를 위해, 우선 물리학 관련 학과 운영 여부와 관계없이 집계한 국내 전체 대학교 중 사립대학교의 비중은 1980년의 76.47\%에서 2008년의 85.71\%까지 증가추세로 나타나다, 이후 폐교 및 통폐합으로 인하여 2023년 81.57\%까지 감소 추세가 나타난다\cite{KESStimeline}. 이와 비교하여 물리학 학부과정을 운영하는 학교 중 사립대학교의 비중은 1980년 75.61\%에서 2008년 66.20\%, 이후 2023년 56.00\%로 사립대학교 소속 물리학 계열 학과의 계열전환 및 모집중단으로 인하여 동 시기 전체 사립대학교 비중 감소보다 더 큰 값으로 비중이 낮아지는 것을 확인할 수 있다. 참고로 Table~\ref{table:classification}의 66.67\%가 1915$\sim$2023년까지의 모든 데이터에 대한 것이므로, 2023년에 남아 있는 물리학 계열 학과 운영 중인 학과 비율 중 사립대학교의 비중이 훨씬 더 낮다는 것은 그만큼 사립대학교에서 운영하는 물리학 계열 학과 수의 급격한 감소가 일어나고 있다는 뜻이다. 또한 2000년 이후 학부과정 신입생 모집을 시작한 3개의 학교는 광주과학기술원(GIST), 대구경북과학기술원(DGIST), 울산과학기술원(UNIST)으로, 이 세 개의 과학기술 특성화 대학에서는 물리학 전공 대학원 과정만을 운영하다 GIST는 2006년에 설치된 ``광과학기술학제학부''를 모태로 하여 2012년 물리학 학부 과정, DGIST와 UNIST는 2014학년도 물리학 학부 과정 모집을 처음 시작한 것을 확인하였다. 

지역과 설립 주체 외에, 대학의 학문 조직 구성으로서의 학과의 위치를 살펴보기 위해 학과가 소속한 단과대학별로 구분해서 나타낸 것이 Fig.~\ref{figure : Fig-3}(c)이다. 1980년 이전에는 과반수의 연구대상이 ``문리대학'' 소속으로 나타났지만, 1980년 이후는 과반수 이상의 연구대상이 ``자연과학대학'' 소속으로 나타났다. 자연과학을 공학이 아닌 인문학과 묶는 `문리대학'의 전통이 사라지면서 생긴 변화로 보인다. 마지막으로, 학과에서 표방하는 정체성을 확인할 수 있는 학과의 이름에 드러나는 특성에 따라 분류한 것이 Fig.~\ref{figure : Fig-3}(d)이다. 검은 선으로는 집계 대상 전체 학과\footnote{여기에만 속한 학과들의 예시: 자연과학대학 반도체물리학과,  과학기술대학 응용물리학과.}, 빨간 선으로는 기초자연과학 키워드\footnote{기초, 수학, 화학, 생명과학, 지구과학, 천문학, 물리학.}로만 구성된 물리학 학부 과정 학과\footnote{예시: 기초과학부 물리학전공, 물리천문학과, 물리화학부 물리학전공, 물리학과, 수물학과, 이학과.}, 그리고 파란 선으로는 ``물리학과'' 및 상위 학부에 포함되지 않은 ``물리학전공''학과\footnote{예시: 물리학과, 이과대학 물리학전공, 자연과학대학 물리학전공.}의 시간에 따른 증감을 나타냈다. 

특히 두드러지는 것이 1990년대 후반 학부제 시행의 결과로 보이는, 독립적으로 운영되는 ``물리학과'' 및 ``물리학전공''의 수가 크게 줄어든 것이며, 최근 교육부에서 각 대학에 권장하고 있는 정책인 `모집 단위 광역화'\cite{news}를 통해 이 현상이 반복될 수도 있음을 추론할 수 있다. 무엇보다, 최근에 전반적인 학과 수가 감소하고 있음에도 불구하고 `반도체'나 `데이터'와 같은 접두사를 붙여서 특성화를 시도한 학과들(검은 선에만 집계됨)과 `물리학과'와 같은 전통적 이름 또는 기초자연과학 키워드만 포함한 학과들 수의 차이는 유지되고 있다. 즉, 명칭 변경으로 특성화된 학과 수의 상대적 비율은 점점 높아지고 있는 것이다. 다시 말해서, 학과 수의 전반적인 감소와 더불어 남은 학과들 중에는 특성화된 학과 비율의 증가가 일어나는 현상이 일어나고 있다는 뜻이다.

\section{인구밀도 대비 지역별 분포 분석}
\label{section : dept scaling}

\subsection{스케일링 분석}
\label{sec:scaling_method}

\begin{figure*}
\includegraphics[width=0.85\textwidth]{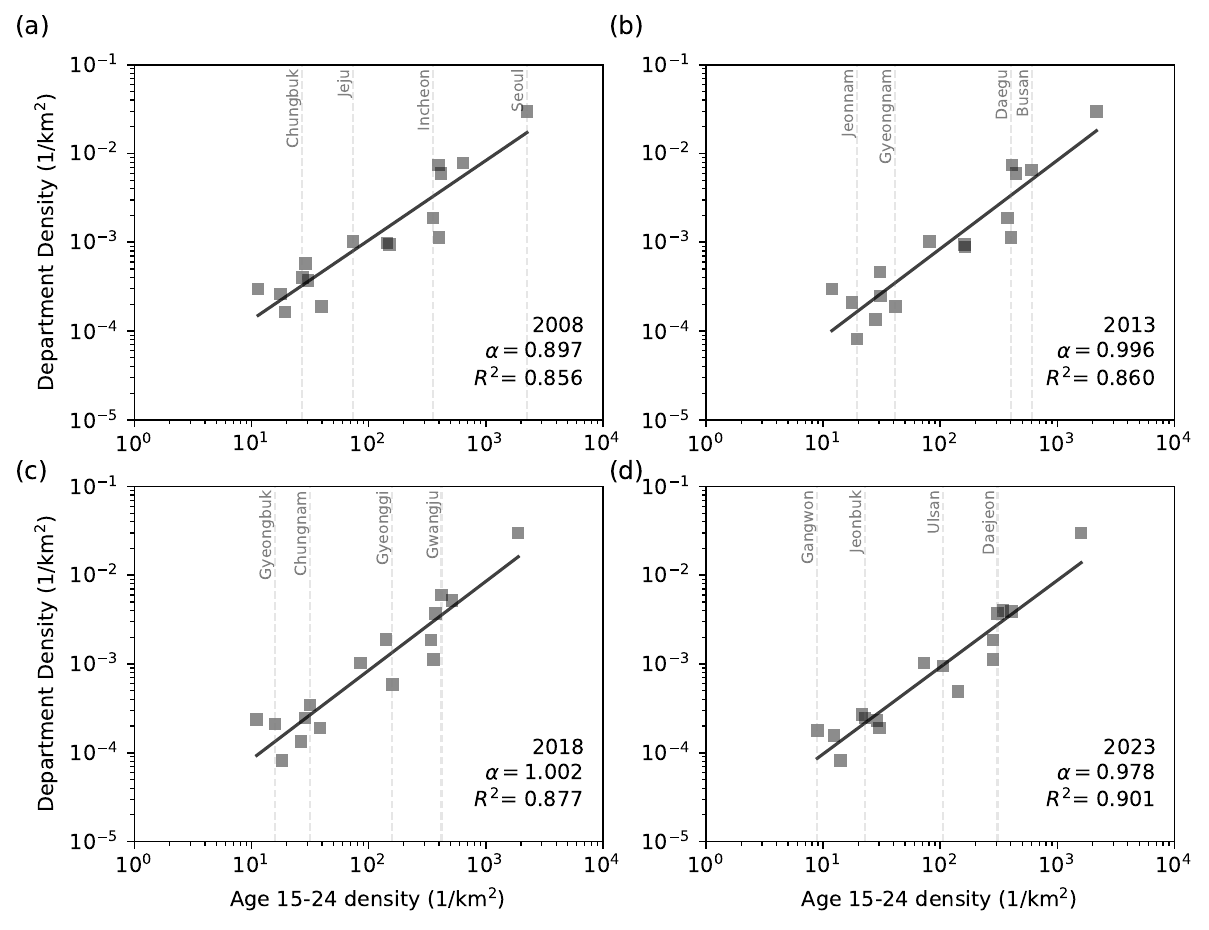}
\caption{The scatter plots show the physics-related-department density against the population density of 15--24-year-olds for each region, along with the linear-regression lines (in the log-log scale), in (a) 2008, (b) 2013, (c) 2018, and (d) 2023. We indicate all of the $16$ administrative regions across the four panels, which are identifiable because each point corresponding to each region shares the same value of the horizontal axis in all of the panels.
}
\label{figure : Fig-4}
\end{figure*}


\begin{figure}
\includegraphics[width=0.5\textwidth]{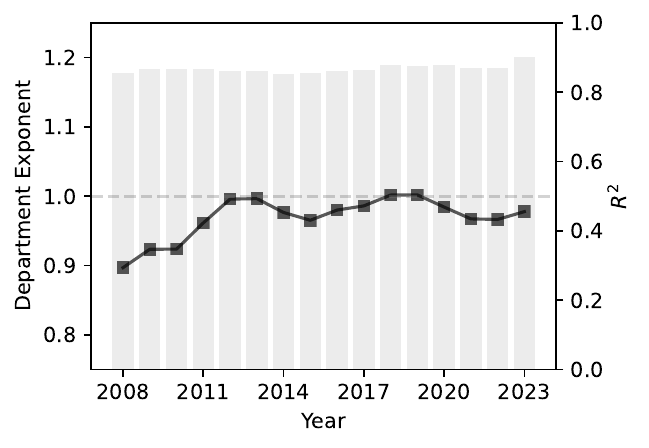}
\caption{The scaling exponent for the physics-related-department density with respect to the population density of 15--24-year-olds (the filled squares---the lines connecting them are the guide to the eyes) and $R^2$ of each regression (the shaded bar chart behind each square) for each year, obtained from the linear regression in the log-log scale.}
\label{figure : Fig-5}
\end{figure}

\begin{figure*}
\includegraphics[width=0.85\textwidth]{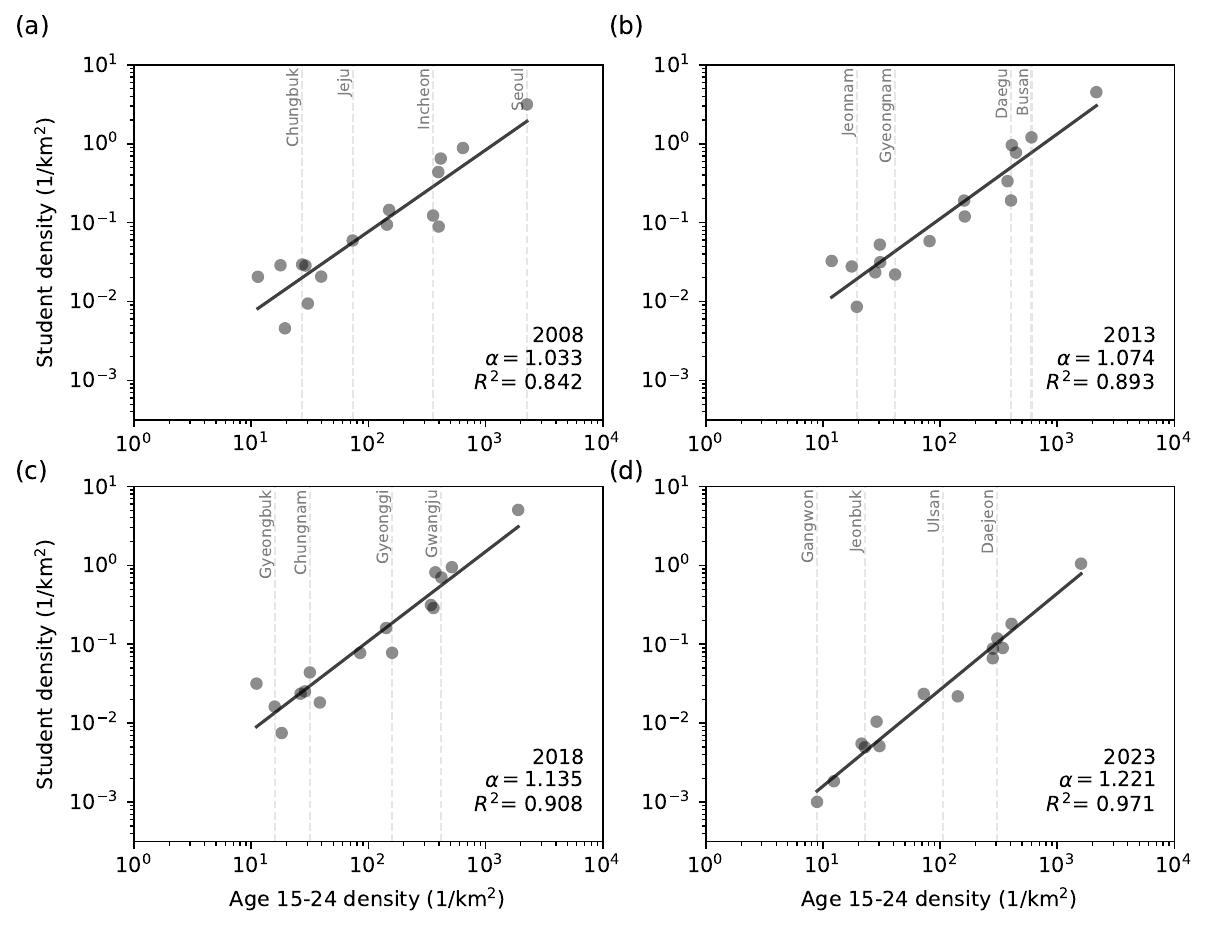}
\caption{The scatter plots show the physics-undergraduate-student density against the population density of 15--24-year-olds for each region, along with the linear-regression lines (in the log-log scale), in (a) 2008, (b) 2013, (c) 2018, and (d) 2023. We indicate all of the $16$ administrative regions across the four panels, which are identifiable because each point corresponding to each region shares the same value of the horizontal axis in all of the panels. 
}
\label{figure : Fig-6}
\end{figure*}


\label{section : stu data}
\begin{figure}
\includegraphics[width=0.5\textwidth]{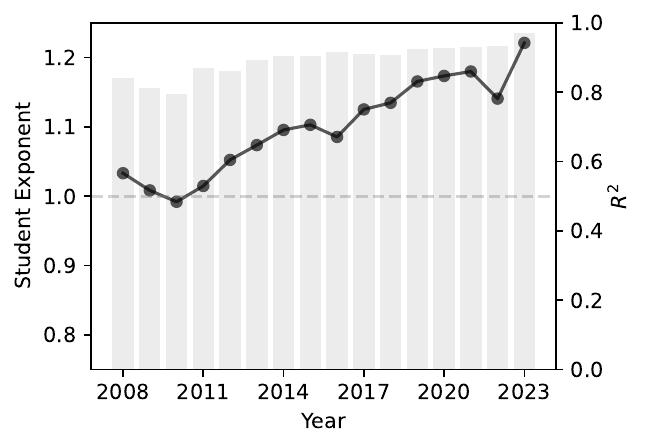}
\caption{The scaling exponent for the physics-undergraduate-student density with respect to population density of 15--24-year-olds (the filled circles---the lines connecting them are the guide to the eyes) and $R^2$ of each regression (the shaded bar chart behind each circle) for each year, obtained from the linear regression in the log-log scale.}
\label{figure : Fig-7}
\end{figure}

\ref{sec:data}절의 데이터에 기반하여, 통계물리학의 스케일링 이론으로 학령인구 감소 및 수도권 대학 선호에 따른 최근 추세를 정량화하였다. 스케일링 관계\cite{PNASScaling} 
\begin{equation}
D =C\rho^{\alpha}
\label{eq:scaling_relation}
\end{equation}
에 기반하여, 로그-로그 스케일에서의 선형 회귀를 식 
\begin{equation}
\log D = \alpha \log \rho + \log C 
\label{eq:scaling_relation_log}
\end{equation}
에 대해 수행하였다. 기존 연구\cite{newman,PNASScaling}에 의하면 지역별 시설 밀도를 $D$, 인구 밀도를 $C$라고 할 때, 각 시설을 이용하는 인구 수를 공평하게 만들도록 최적화(주로 상업 시설)하면 식~\eqref{eq:scaling_relation}에서 $\alpha = 1$인 선형(linear) 관계를 따르게 되고 시설에 대한 접근성의 총합을 공평하게 만들도록 최적화(주로 공공 시설)하면 $\alpha = 2/3$인 준선형(sublinear)관계를 따르게 된다. 본 연구에서 $D$는 지역별 물리학 학부 과정 학과 및 재학생 밀도, $\rho$는 고등교육 대상 인구 밀도, $C$는 회귀 분석을 위하여 사용되는 임의의 상수이다. 선형 회귀를 통해, 각 데이터에 수행한 회귀분석을 통하여 스케일링 지수 $\alpha$와 결정계수 $R^2$을 도출하고 그 의미를 논할 것이다.

\subsection{지역별 물리학 관련 학과 분포}
\ref{sec:data}절에서 소개한 데이터에 기반하여 지역별 학과의 수를 지역 면적으로 나눈 학과 밀도에 로그를 취한 값, 지역별 고등 교육 대상 연령 밀도에 로그를 취한 값을 통하여 선형 회귀 분석을 수행하였다. 로그-로그 스케일로 그린 Fig.~\ref{figure : Fig-4}에서 각 점의 가로축 값은 지역별 고등 교육 대상 연령의 밀도, 세로축 값은 지역별 물리학 학부 과정 학과 밀도를 의미한다. 2008년부터 2023년까지 분석을 시행했으며 대표적으로 2008년, 2013년, 2018년, 2023년의 값을 산포도(scatter plot)로 제시하였다. 산포도를 가로지르는 직선은 이 데이터를 기반으로 선형 회귀 분석을 수행한 결과로, 기울기가 식~\eqref{eq:scaling_relation_log}의 스케일링 지수 $\alpha$와 같으며, 선형 회귀 분석에서 제시한 모델과 실제 데이터와의 일치 정도를 나타내는 결정계수(coefficient of determination) $R^2$이 모든 연도에서 $0.85$ 이상으로 도출되어 스케일링 지수 값들의 통계적 유의미함을 알려준다.

앞서 언급한 선형 회귀 분석을 통하여 연도별 지역별 물리학 학부 과정 학과의 스케일링 지수(왼쪽 세로축)를 사각형 점을 연결한 꺾은선 그래프, 연도별 $R^2$의 값(오른쪽 세로축)을 뒤쪽의 옅은 회색 막대그래프로 나타낸 그림을 Fig.~\ref{figure : Fig-5}에 나타냈다. 각 값은 $0.89$에서 $1.00$의 범위로, 기존 연구\cite{PNASScaling}에서 밝힌 대한민국 지역별 전 연령 인구 밀도와 대학교 밀도의 스케일링 지수인 $0.93$과 유사한 값으로 나타났다. 그러나 학과 스케일링 지수는 시간이 흐름에 따라 준선형에서 선형으로 증가하는 경향이 나타났다. 이것은 고등교육 대상 인구 밀도가 낮은 지역에서는, 공평한 접근성을 위해 필요한 물리학 학부 과정 학과의 비율보다 실제 학과의 수가 점점 더 적어진다는 뜻이다. 선행 연구\cite{OurSaemulli}에서 조사한 다른 학과들의 사례와 마찬가지로, 예전에는 지역별 접근성을 고려하여 교육 시설의 밀도와 인구 밀도 간의 차이가 상대적으로 적었으나 최근들어 인구 밀도에 맞추어 지역별로 차이가 심해지는 경향이 있음을 알 수 있다. 

\subsection{지역별 물리학 전공 학부 재학생 분포}
\ref{sec:data}절의 물리학 학부 과정 학과 데이터를 통하여, KESS, 학교/학과별 데이터\cite{kessdata}
에 기반한 분석을 수행하였다. 매년 달라지는 현행 ``물리학'' 학부 과정 운영 학과 데이터를 선별하고, 각 학과에 소속된 학생 데이터에 \ref{sec:scaling_method}절에서와 같은 방법을 사용하여 선형 회귀 분석을 수행하였다. 역시 로그-로그 스케일로 그린 Fig.~\ref{figure : Fig-6}에 표시된 각 사각형 점의 가로축 값은 지역별 고등 교육 대상 연령의 밀도, 세로축 값은 지역별 물리학 학부 과정 재학생 밀도를 의미한다. \ref{sec:scaling_method}절에서와 같은 방법으로 2008년부터 2023년까지 분석을 시행했으며 대표적으로 2008년, 2013년, 2018년, 2023년의 산포도와 선형 회귀 결과식을 나타냈다. 

선형 회귀 분석 결과를 정리해서 나타낸 Fig.~\ref{figure : Fig-7}\은, 지역별 물리학 학부 과정 재학생의 스케일링 지수(왼쪽 세로축)를 원형 점을 연결한 꺾은선 그래프, 연도별 $R^2$의 값(오른쪽 세로축)을 뒤쪽의 옅은 회색 막대그래프로 나타낸 그림이다. $R^2$은 입학 후 일정 학기 후 전공이 결정되는 학부제가 주로 운영된 2010년대 초반을 제외하면, 집계 기간에 $0.8$ 이상의 높은 값으로 나타났다. 학생 스케일링 지수 값은 0.99에서 1.22의 범위로 선형에서 초선형(superlinear)으로 증가하는 추세를 나타났다. 이러한 초선형 값은 기존 연구\cite{PNASScaling}에서 은행, 유료주차장과 같이 이윤 추구 시설의 최적값보다 더 심하게 인구밀도가 높은 곳으로 시설이 몰려 있는 상황에 해당한다. 즉 이러한 높은 학생 스케일링 지수는, 접근성에 대한 공정함을 어느 정도 포기하며 인구 밀도에 맞춰 분포하는 상황보다 더 심하게 인구밀도가 높은 곳으로 학생 정원이 점점 더 이동하고 있다는 뜻이며, 학과의 소재 지역에 따른 물리학 학부 과정 재학생 수의 양극화가 학과 수의 양극화보다 더 심하게 발생하고 있다는 의미이다.  

\section{결론}
본 연구에서는, 국내 대학 물리학 학부 과정 운영에 대한 데이터를 충실하게 수집하여 시간에 따른 변화 및 고등교육 대상 인구 밀도 대비 지역별 분포를 분석하였다. 분석 결과 2000년 이후 급격하게 전반적인 감소가 일어나고 있으며, 서울 이외의 지역에 소재한 대학교 소속의 변화, 그 중에서도 특히 지역 소재 사립 대학교에서의 감소가 두드러짐을 확인하였다. 구체적으로 서울 소재 학교의 학과와 학과의 정원은 2010년대 초반과 비교하여 비교적 일정한 수치를 유지하는 방면, 지역 소재의 학교들은 학과의 수가 줄어들 뿐만 아니라, 지속해서 신입생을 모집하고 있는 학과라도 입학생 정원 및, 재학생 수가 줄어드는 것을 확인하였다. 학과의 명칭에서도 전체 비율에 비해 `반도체'나 `전자'와 같은 이름을 추가해서 특성화를 통한 차별화를 시도하는 학과들의 비율이 증가하고 있음을 알 수 있다. 또한, 고등교육 대상 인구 밀도 대비 지역별 분포로부터, 수도권 집중화가 학과 밀도와 재학생 밀도 모두에서 증가하고 있으며 스케일링 지수로 볼 때 후자가 더 심하게 나타남을 확인하였다.

기존 연구\cite{OurSaemulli}에서 지역별 학령인구 밀도와 ``수학$\cdot$물리$\cdot$천문$\cdot$지리'' 및 ``생물$\cdot$화학$\cdot$환경''의 자연과학(기초)계열에 대하여 수행한 스케일링 분석 연구를 이 결과와 비교 분석할 수 있다. 인구 밀도의 대상이 완전히 같은 대상이 아님을 고려해도, 물리학 학부 과정 학과 및 학생의 각 스케일링 지수가 자연과학(기초)보다 큰 것을 확인할 수 있다. 자연과학(기초)의 학과 및 학생의 스케일링 지수가 준선형 값을 유지하는 것에 반해, 물리학 학부 과정의 학과는 준선형에서 선형, 그리고 물리학 학부 과정 학생은 선형에서 초선형으로 증가하는 추세가 나타났다. 이는 고등교육 대상 인구의 감소를 고려하여도 지역에 따른 양극화가 심화되고 있음을 의미한다. 교육뿐만이 아니라, 이러한 전반적인 학과 및 학생의 감소와 양극화는 연구에 있어서도 영향을 끼칠 것으로 예상된다. 연구 그룹의 규모에 따른 생산량의 상관관계를 분석한 연구\cite{SZhang20252}에 따르면 그룹의 규모, 특히 소속 연구원의 수와 연구생산성이 양의 상관관계로 나타난다. 이를 고려하여 물리학 후학 양성과 더불어 신규 전공자의 합류를 통한 연구의 지속적인 수행 및 발전을 위해서 학회차원에서 현재의 학과 및 학생감소 및 양극화 현황을 더욱 예의주시할 필요가 있다. 

Appendix~\ref{appendix : data}에 우리가 수집하고 정제한 데이터를 공개했음을 다시 한 번 알리고, 이를 통해 다른 연구자들도 이러한 정량적인 분석을 다각도로 할 수 있기를, 고등교육 및 연구 정책 수립에 그러한 분석 결과가 유의미하게 사용될 수 있기를 기원한다. 추가 연구에 대한 제안으로서, 학생들의 생애 주기 전반을 고려하기 위한 지역별 일자리 분포와 같은 다른 사회경제적 지표와의 비교 및 학과를 졸업한 학생이 일자리를 찾아 이주하는 모형 연구\cite{Simini2012} 등을 생각할 수 있으며, 물리학 외의 다른 과학기술계열(science, technology, engineering, and mathematics: STEM) 분야와의 좀 더 정밀한 비교 역시 의미 있는 연구가 될 것이다. 특히 미국에서 2011년 대비 2021년 기초과학을 포함한 STEM 전공자 수와 비율이 증가한 사례\cite{stemnews}와 비교하여 물리학을 포함하여 기초과학 전공별 학생 거듭제곱 지수 증감에 대해서 변인 분석 연구를 수행하는 연구 등이 가능할 것이다. 

\section*{감사의 글}
이 논문은 한국연구재단 (NRF-2021R1C1C1004132, NRF-2022R1A4A1030660)의 예산지원으로 수행된 연구입니다.
\newline

\appendix

 \section*{Appendix}
\subsection{원본 데이터}
\label{appendix : data}
\begin{itemize}
    \item\href{https://docs.google.com/spreadsheets/d/1xaQgcpUiWgIWK-TKu_XBQsfWBtVt2F_Z8rA_nEOsZFg/edit?usp=sharing}{학과데이터(1915-2023).xlsx}
        \begin{itemize}
        \item 집계 대상의 2023학년도 기준 소재 지역, 설립 주체, 본·분교 여부 그리고 1915년부터 2023년도까지의 명칭, 소속 변경을 수집 및 정리하였다. 학부 신입생 모집 기준으로 학과 개설 및 첫 입학생 모집(빨간색), 명칭 혹은 소속 변경(초록색), 그리고 학부 신입생 모집 중단(파란색)으로 구분하였다.
        \end{itemize}
    \item\href{https://docs.google.com/spreadsheets/d/1ox2BAip7HRud3kRi1MjrszFySnLULNsF1zi-unPj4gU/edit?usp=sharing}{학생데이터(1990-2023).xlsx}
        \begin{itemize}
        \item 교육부의 교육통계연보를 기반으로 하여 집계 대상의 1990학년도부터 2023학년도까지의 봄학기 기준으로 전체 학부생(총계), 이공계열 학부생(이공계열), 자연과학 계열 학부생(자연과학 계열) 및 전공 미확정자를 제외한 물리학 학부 과정 학과 소속 학부생 수를 취합하여 정리하였다. 또한 세부 데이터가 공개된 2008학년도부터는 학교/학과별 데이터셋에서의 Section \ref{section : dept data}에 해당하는 학과를 선별하여 정리하였다.  
        \end{itemize}
\end{itemize}

\subsection{예외대상}
 \label{appendix : exception}
	\begin{itemize}
		\item 추가대상
 \begin{itemize}
\item 학부 과정 학생 전원 공통학부 소속 특수성 고려 
 \begin{itemize}
\item DGIST 융복합대학 기초학부 \newline 기초과학 물리학트랙
\end{itemize}
\end{itemize}
 \begin{itemize}
\item 인접학문 학과와의 통합으로 인한 \newline교육부 집계 예외 사례
 \begin{itemize}
\item 강릉원주대학교 자연과학대학 \newline수학물리학부 물리$\cdot$에너지전공
\item 세종대학교 자연과학대학 물리천문학과

\end{itemize}

\end{itemize}
		\item 제외대상
\begin{itemize}
\item 교차검증 후 대학알리미 집계 오류 판단
 \begin{itemize}
\item 신라대학교 공과대학 나노생체화학공학부
 \item 인제대학교 \newline소프트웨어대학 드론IoT시뮬레이션학부, \newline AI융합대학 드론IoT시뮬레이션학부
 \end{itemize}
\item 융합전공 학과
 \begin{itemize}
\item 충북대학교 바이오헬스공유대학 \newline방사광융합학과 
 \item 서울시립대학교 융합전공학부 \newline물리학-전자물리학, \newline물리학-나노반도체물리학
 \end{itemize}
\end{itemize}
	\end{itemize}

\subsection{세부 지역별 물리학 관련 학과 수의 증감}

\begin{figure*}
\includegraphics[width=0.85\textwidth]{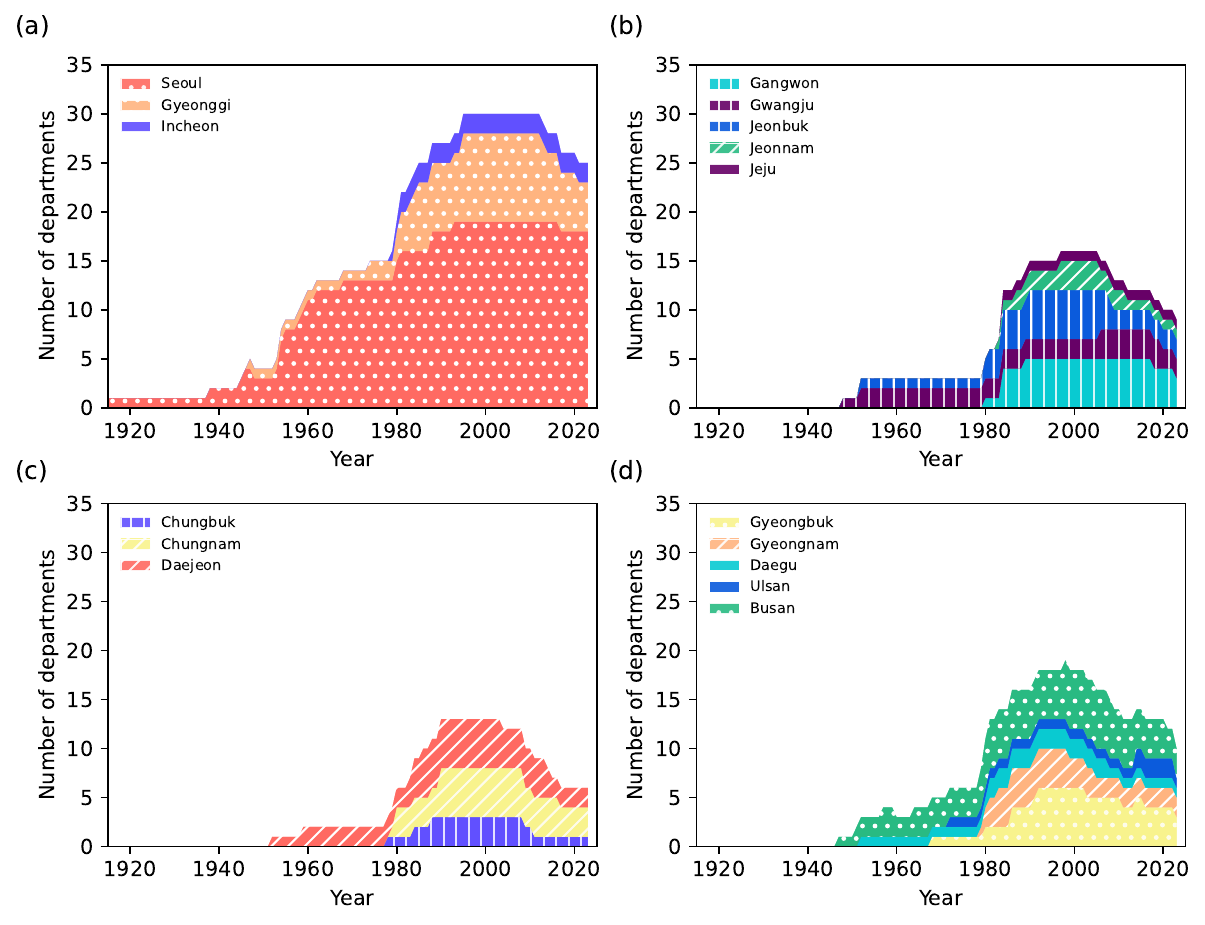}

\caption{The temporal change of the number of physics-related departments, which are categorized by administrative regions classified into four regions: (a) the capital region, (b) the Gangwon, Jeolla, and Jeju regions, (c) the Chungcheong region, and (d) the Gyeongsang region. For each panel, we plot the cumulative data stacked on top of one another as in Figs.~\ref{figure : Fig-3}(b) and (c).}
\label{figure : Fig-appendix}

\end{figure*}

Figure~\ref{figure : Fig-appendix}에서, Fig.~\ref{figure : Fig-3}(a)에 표시한 지역별 물리학 관련 학과 수의 증감을 광역자치단체 수준으로 구분하여 자세히 나타내었다.

\begin{references}
\bibitem{studecline} S.\,J. Ban, The Journal of Educational Research (교육종합연구) \textbf{14}, 213 (2016). 

\bibitem{Univalram} Higher Education in Korea (대학알리미) 2024년 상반기 기준 학교별 학부/과(전공) 리스트(2024.03.06.기준), \href{https://academyinfo.go.kr/brd/brd0520/selectDetail.do?ntce_sntc_sno=151&bbs_gubun=rfbr&no=147#none}{https://academyinfo.go.kr/index.do} (accessed March 26, 2024).

\bibitem{kpsUniv} The Korean Physical Society (한국물리학회) 물리학정보 유용한사이트 대학물리학과, \href{https://www.kps.or.kr/content/def/view.php?ft=64}{https://www.kps.or.kr/content/def/view.php?ft=64} (accessed March 26, 2024).

\bibitem{newman} M.\,T. Gastner and M.\,E.\,J. Newman,
    \refdoi{Phys. Rev. E \textbf{74}, 016117 (2006)}{10.1103/PhysRevE.74.016117}

\bibitem{PNASScaling} J. Um, S.-W. Son, S.-I. Lee, H. Jeong, and B.\,J. Kim, 
    \refdoi{Proc. Natl. Acad. Sci. U.S.A. \textbf{106}, 14236 (2009)}{10.1073/pnas.0901898106}

\bibitem{OurSaemulli} J. Seo, G. Gim, S. Kim, M. Ha, and S.\,H. Lee,
    \refdoi{New Phys.: Sae Mulli, \textbf{73}, 858 (2023)}{10.3938/NPSM.73.858}

\bibitem{kosis} KOSIS (KOrean Statistical Information Service, 국가통계포털), \newline
    \href{https://kosis.kr/index/index.do}{https://kosis.kr/index/index.do} (accessed March 26, 2024).

\bibitem{kps50th} The 50-year History of the Korean Physical Society (한국물리학회 50년사) \newline \href{https://www.kps.or.kr/content/50years/html/kps195.htm}{https://www.kps.or.kr/content/50years/html/kps195.htm} (accessed March 26, 2024).

\bibitem{KESStimeline} KESS (Korean Education Statistic Service, 교육통계서비스), 시계열 통계 대학 설립별 학교수 (1980--2023), \href{https://kess.kedi.re.kr/kessTheme/timeStats}{https://kess.kedi.re.kr/kessTheme/timeStats} (accessed March 26, 2024).

\bibitem{news} Ministry of Education, [설명자료] 교육부는 충실한 의견수렴 결과를 반영하여 2024년 대학혁신지원 사업계획을 확정할 예정입니다. \href{https://www.moe.go.kr/boardCnts/viewRenew.do?boardID=295&boardSeq=97592&lev=0&searchType=null&statusYN=W&page=1&s=moe&m=020401&opType=N}{https://www.moe.go.kr/boardCnts/listRenew.do} (accessed March 26, 2024)

\bibitem{kessdata} KESS (Korean Education Statistic Service, 교육통계서비스), 학교/학과별 데이터셋, \href{https://kess.kedi.re.kr/contents/dataset}{https://kess.kedi.re.kr/contents/dataset} (accessed March 26, 2024).

\bibitem{SZhang20252} S. Zhang, K.\,H. Wapman, D.\,B. Larremore, and A. Clauset, \refdoi{Sci. Adv. \textbf{8}, eabq7056 (2022)}{10.1126/sciadv.abq7056}

\bibitem{Simini2012} F. Simini, M.\,C. Gonz{\'a}lez, A. Maritan, and A.-L. Barab{\'a}si, \refdoi{Nature {\bf 484}, 96 (2012)}{doi.org/10.1038/nature10856}

\bibitem{stemnews} A. Van Dam (2022, September 2). The most-regretted (and lowest-paying) college majors. \textit{The Washington Post}. \href{https://www.washingtonpost.com/business/2022/09/02/college-major-regrets/}{https://www.washingtonpost.com/business/2022/09/02/college-major-regrets/} (accessed March 26, 2024)


\end{references}
\end{document}